\documentclass[10pt,final,doublecolumn]{IEEEtran}
\IEEEoverridecommandlockouts
\usepackage{amsmath}
\usepackage{amssymb}
\usepackage{latexsym}
\usepackage{graphicx}
\usepackage{bbding}
\usepackage{indentfirst}
\usepackage{cases}
\usepackage{supertabular}
\usepackage{algorithm,algorithmic}
\usepackage{subeqnarray}
\usepackage{color}
\usepackage{bm}
\usepackage{stfloats}
\allowdisplaybreaks[4]

\begin{document}
\title{Integrated Sensing, Computation and Communication in B5G Cellular Internet of Things}
\author{Qiao Qi, Xiaoming Chen, Caijun Zhong, and Zhaoyang Zhang
\thanks{Qiao Qi ({\tt qiqiao1996@zju.edu.cn}), Xiaoming Chen ({\tt chen\_xiaoming@zju.edu.cn}), Caijun Zhong ({\tt caijunzhong@zju.edu.cn}) and Zhaoyang Zhang ({\tt ning\_ming@zju.edu.cn}) are with the College of Information Science and Electronic Engineering, Zhejiang University, Hangzhou 310027, China.}}\maketitle

\begin{abstract}
In this paper, we investigate the issue of integrated sensing, computation and communication (SCC) in beyond fifth-generation (B5G) cellular internet of things (IoT) networks. According to the characteristics of B5G cellular IoT, a comprehensive design framework integrating SCC is put forward for massive IoT. For sensing, highly accurate sensed information at IoT devices are sent to the base station (BS) by using non-orthogonal communication over wireless multiple access channels. Meanwhile, for computation, a novel technique, namely over-the-air computation (AirComp), is adopted to substantially reduce the latency of massive data aggregation via exploiting the superposition property of wireless multiple access channels. To coordinate the co-channel interference for enhancing the overall performance of B5G cellular IoT integrating SCC, two joint beamforming design algorithms are proposed from the perspectives of the computation error minimization and the weighted sum-rate maximization, respectively. Finally, extensive simulation results validate the effectiveness of the proposed algorithms for B5G cellular IoT over the baseline ones.
\end{abstract}


\section{Introduction}
Nowadays, the rapid development of internet of things (IoT) incurs the exponential growth of terminal devices and the surge of data traffic. It is predicted that over 75.4 billion devices will be linked to the internet all over the world by 2025, which means a roughly $400\%$ growth for the ten years compare to 15.4 billion in 2015 \cite{5G1,5G0}. In this context, 3GPP have launched the fifth-generation (5G) cellular IoT in 2015 \cite{3GPP}, so as to support various new applications, e.g., virtual reality (VR), augmented reality (AR), autonomous driving and etc \cite{5G2,5G3}. In general, these applications require ultra-high accuracy of sensing, ultra-low latency of computation, and ultra-high speed of communication among a massive number of IoT devices. However, massive machine type of communication (mMTC) for 5G cellular IoT only emphasizes the number of connections, but does not demand real-time, reliability and high-speed. Therefore, it is desired to design beyond 5G (B5G) cellular IoT networks with distinct service provisions.

In the era of IoT, most devices are used for environment sensing. For instance, there are a large number of sensors for temperature measure and cameras for video capture in the city. Especially, with the development of autonomous driving, every car will be equipped with numerous sensors to sense surroundings. As a result, there are a massive column of sensing information that has to be transferred from IoT devices to the base station (BS) \cite{Massive}. However, it is not a trivial task to transfer highly accurate sensing information over limited radio spectrum. Specifically, the accuracy of sensing information is mainly determined by the number of quantization bits. Due to limited radio spectrum, traditional orthogonal multiple access (OMA) schemes cannot support high-capacity transmission of a massive number of IoT devices. To solve this challenge, non-orthogonal multiple access (NOMA) is applied into cellular IoT to realize high-speed transmission over limited radio spectrum \cite{NOMA0}-\cite{NOMA1}. The authors in \cite{NOMA2} studied a uplink millimeter wave massive system with NOMA, and proposed a power allocation algorithm to maximize the energy efficiency. Moreover, a grant-free NOMA scheme was designed to enhance the performance of the uplink system with massive access in \cite{NOMA3}.

\begin{figure}
 \centering
\includegraphics [width=0.5\textwidth] {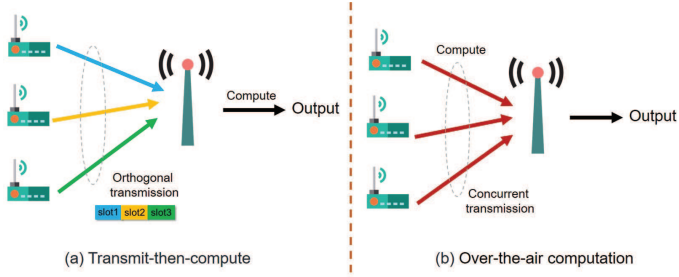}
\caption {Comparison of conventional computation and AirComp.}
\label{FigAircomp}
\end{figure}

On the other hand, stimulated by the demands of fast data aggregation for IoT scenarios, B5G cellular IoT is converting from a data-centric network to a computation-centric one. The advanced information processing technologies, such as artificial intelligence (AI) and data mining, will provide ubiquitous computing and intelligent services to effectively realize analysis and processing of massive data from IoT devices, which means that B5G cellular IoT may be more concerned about the computation results of the data, e.g., the sum, the maximum, the minimum and etc, rather than the individual data itself. For instance, an IoT-based humidity monitoring system is only interested in the average of humidity in certain region, instead of collecting all observations from sensors. For realizing massive data computation from IoT devices, the conventional approach of \emph{transmit-then-compute} is no longer applicable for B5G cellular IoT due to the excessively high latency and the low spectrum efficiency. To address this issue, a promising solution called \emph{over-the-air computation} (AirComp) has been proposed and raised wide interests \cite{AirComp1,AirComp2}, which exploits the superposition property of wireless multiple access channels (MACs) to compute a class of \emph{nomographic functions} \cite{Nomofun} of distributed data from IoT devices via concurrent transmission, c.f., Fig \ref{FigAircomp}. More importantly, the accuracy of computation enabled by AirComp can be improved as the number of simultaneous IoT devices increases. Compared to the approach of transmit-then-compute, AirComp can significantly decrease the data aggregation latency by a factor equal to the number of IoT devices. In fact, the notion of AirComp originated from information theory. The author in \cite{AirComp01} studied the issue of computing functions over MACs, and proposed a coding technique for reliable distributed computations by utilizing the interference resulted from simultaneous transmissions. Then, it was found that if the transmitted data is an independent and identically distributed (i.i.d.) Gaussian  random variable, a simple analog transmission can achieve the minimum distortion even without coding \cite{AirComp02}, which motivated a series of works on the implementation of AirComp \cite{AirComp03}-\cite{AirComp05}.

\subsection{Related Works and Motivation}
Compared to AirComp in pioneering works only focusing on scalar-function computation, AirComp in B5G IoT networks can spatially multiplex multi-function computation by exploiting spatial degrees of freedom provided by the large-scale antenna array at the BS, namely MIMO AirComp \cite{AirComp4}-\cite{AirComp6}. In \cite{AirComp4}, the authors studied the MIMO AirComp with multiple linear functions of Gaussian sources by using antenna arrays. In \cite{AirComp5} and \cite{AirComp5.5}, the integration of energy supply and data aggregation was investigated for wirelessly powered MIMO AirComp systems.  The authors in \cite{AirComp6} designed a reduced-dimension MIMO AirComp framework for clustered IoT networks.

AirComp have received considerable attention, since it is a promising method to reduce the computation latency in cellular IoT. Yet, AirComp only addresses the issue of computation. As mentioned earlier, B5G cellular IoT usually has multiple tasks, e.g., sensing and computation. In order to realize accurate sensing and computation, B5G cellular IoT has to provide efficient communication for both the sensing signals and the computation signals from a massive number of devices with limited wireless resources. In this context, NOMA techniques has to be adopted for B5G cellular IoT. NOMA leads to severe co-channel interference, especially in the scenario of massive IoT. In fact, co-channel interference has different impacts on the performance of sensing and computation. Specifically, for sensing, the signal stream from each individual device should be separated from the mixed received signal. Thus, co-channel interference decreases the quality of the sensing signal \cite{Bad1,Bad2}. For computation, multiple data streams from different devices are fused at the BS. Hence, co-channel interference can improve the accuracy of computation \cite{Good1,Good2}. In other words, the original harmful interference can be exploited to enhance the performance of computation. In order to depress the impact of co-channel interference on sensing but enhance the impact of co-channel interference on computation, it is desired to utilize transmit and receive beamforming to coordinate the interference. For sensing, there already exist many works about beamforming design \cite{BF1}-\cite{BF3}. Especially in B5G cellular IoT, the BS equipped with a large-scale antenna array has ultra-high spatial degrees of freedom to mitigate co-channel interference \cite{LSantenna1,LSantenna2}. For AirComp, most of the existing works about beamforming design just adopted simple schemes, e.g., zero-forcing beamforming \cite{ZFBF} and uniform-forcing beamforming \cite{simpleAir}.  However, due to the existence of co-channel interference between computation signals and sensing signals, the existing beamforming schemes cannot be applied to the scenario integrating sensing, computation and communication (SCC) directly. Thus, it is necessary to design new transmit and receive beamforming schemes for the integration of SCC in B5G cellular IoT.

\subsection{Contributions}
In this paper, we consider a general B5G cellular IoT network integrating SCC, where transmit and receive beamforming are respectively adopted at the IoT devices and the BS to spatially multiplex multi-function AirComp and multi-stream sensing. The contributions of this paper are three-fold:

\begin{enumerate}

\item We propose a comprehensive design framework for B5G cellular IoT integrating SCC. The originally harmful interference caused by simultaneous transmission is exploited to enhance the overall performance of B5G cellular IoT with multiple tasks.

\item We analyze the impacts of transmit and receive beamforming on the performance of sensing and computation for B5G cellular IoT. Specifically, we select the distortion of computation results measured by mean square error (MSE) and the weighted sum-rate of sensing information as the performance metrics of computation and sensing, respectively.

\item We present two optimization problems, which are respectively formulated from the aspects of minimizing the computation error and maximizing the weighted sum-rate by jointly optimizing transmit and receive beamforming. Then, we provide two low-complexity but effective beamforming design algorithms to improve the overall performance for B5G cellular IoT integrating SCC.
\end{enumerate}

\subsection{Organization and Notations}
The rest of this paper is outlined as follows: Section II gives a concise introduction of a B5G cellular IoT network integrating SCC. Section III focuses on the design of algorithms for transmit and receive beamforming from the perspectives of minimizing the computation error and maximizing the weighted sum-rate, respectively. Section IV presents extensive simulation results to validate the effectiveness of the proposed algorithms. Finally, Section V summarizes this paper.

\emph{Notations}: We use bold upper (lower) letters to denote matrices (column vectors), $(\cdot)^H$ to denote conjugate transpose, $\|\cdot\|_F$ to denote Frobenius norm of a matrix, $\|\cdot\|$ to denote $L_2$-norm of a vector, $|\cdot|$ to denote absolute value, $\mathrm{Re}\{\cdot\}$ to denote the real parts of matrices, $\mathbb{E}\{\cdot\}$ to denote expectation, $\text{tr}(\cdot)$ to denote trace of a matrix, $\text{Rank}(\cdot)$ to denote rank of a matrix,  ${{\mathbb{C}}^{m\times n}}$ to denote the set of $m$-by-$n$ dimensional complex matrix,  ${{\mathbb{R}}^{m\times n}}$ to denote the set of $m$-by-$n$ dimensional real matrix and $\mathcal{CN}(\mu,\sigma^2)$ to denote the circularly symmetric complex Gaussian (CSCG) distribution with mean $\mu$ and variance $\sigma^2$.

\section{System Model}
\begin{figure*}
 \centering
\includegraphics [width=0.9\textwidth] {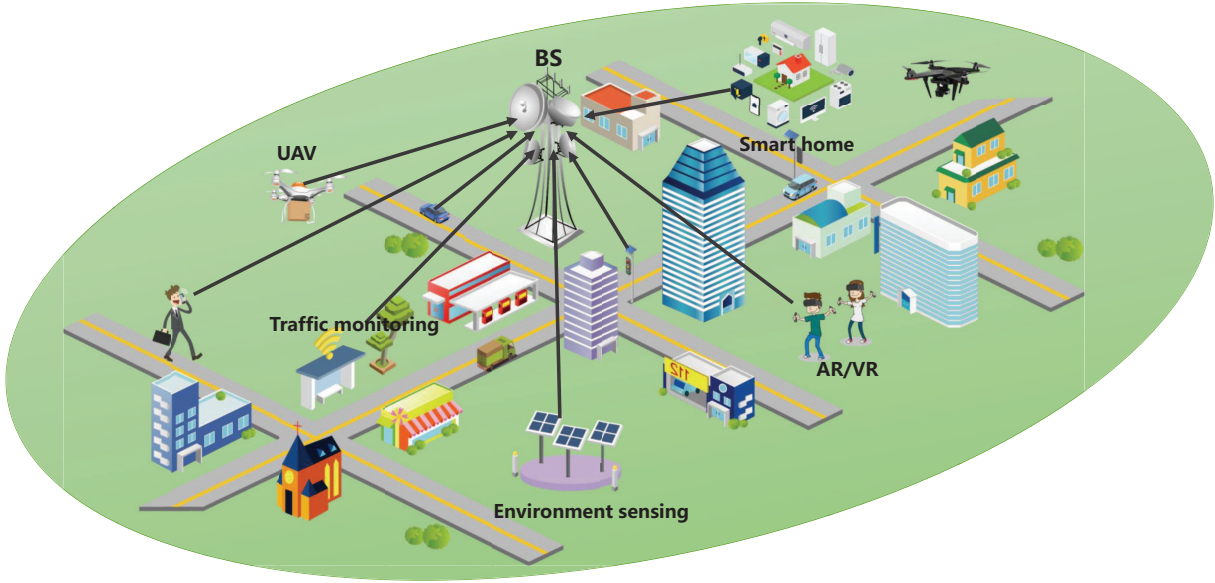}
\caption {A model of B5G cellular IoT network integrating SCC.}
\label{Fig1}
\end{figure*}

\begin{figure*}
 \centering
\includegraphics [width=1.0\textwidth] {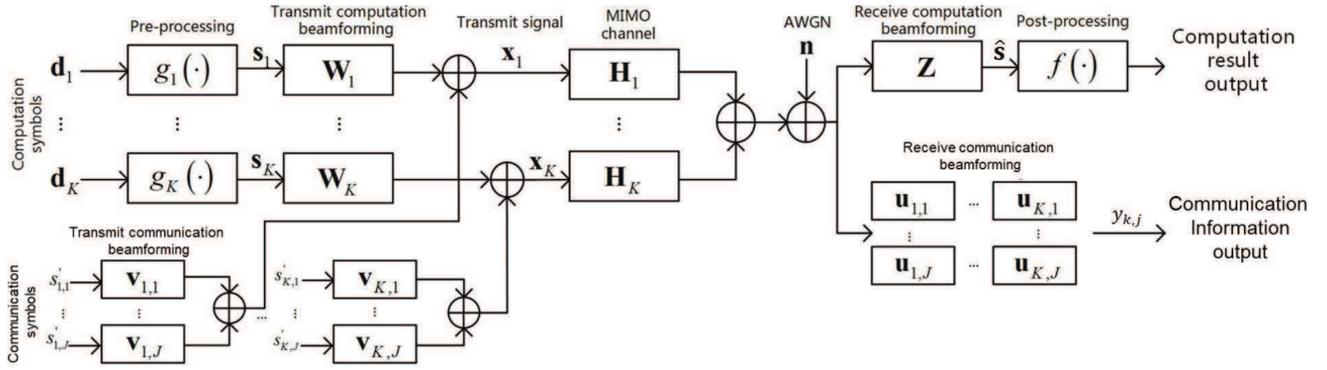}
\caption {The system block diagram for the proposed model.}
\label{Fig2}
\end{figure*}

Let us consider a B5G cellular IoT network comprising a BS equipped with $N$ antennas and $K$ multi-modal IoT user equipments (UEs) equipped with $M$ antennas each, c.f., Fig. \ref{Fig1}. IoT UEs have two fundamental tasks, namely sensing and computation. Specifically, IoT UEs conduct environment sensing and information computation simultaneously, which are fused at the BS by communication. As seen in Fig. \ref{Fig2}, each IoT UE carries out beamforming for coordinating the sensing signal and computation signal to be transmitted respectively, and sends a superposition coded signal to the BS over the uplink channel. On the one hand, by exploiting the superposition property of wireless MACs, the BS receives the computation results directly via concurrent data transmission without recovering individual data, and then utilizes a computation receiver to obtain the targeted function signal. On the other hand, the BS decodes the sensing signals of each UE through the sensing receivers.

\begin{table}[h]
\centering
\caption{Some examples of nomographic functions }\label{Functions}
\begin{tabular}{|c|c|c|c|}
\hline
Functions & $g_k$& $f$& $q$\\ \hline
Arithmetic Mean &$ g_k=d_k$ &$f=1/K$ &  $q=\frac{1}{K}\sum\nolimits_{k=1}^{K}{{{d}_{k}}}$  \\\hline
Weighted Sum& $g_k=\vartheta_kd_k$&$f=1$ &$q=\sum\nolimits_{k=1}^{K}{{{\vartheta }_{k}}{{d}_{k}}}$ \\ \hline
Geometric Mean & $g_k=\ln(d_k)$& $f=\exp(\cdot)$& $q={{\left( \prod\nolimits_{k=1}^{K}{{{d}_{k}}} \right)}^{1/K}}$ \\\hline
Polynomial & $g_k=\vartheta_kd_k^{\beta_k}$ & $f=1$ &$q=\sum\nolimits_{k=1}^{K}{{{\vartheta }_{k}}{{d}_{k}}^{{{\beta }_{k}}}}$ \\\hline
Euclidean Norm &$g_k=d_k^2$ &$f=\sqrt{(\cdot)}$& $q=\sqrt{\sum\nolimits_{k=1}^{K}{d_{k}^{2}}}$\\\hline
\end{tabular}
\end{table}

Without loss of generality, it is assumed that each IoT UE logs data of $L$ heterogeneous parameters to be computed and $J$ heterogeneous parameters sensed from the environment or human, which generate a computation symbol vector ${{\mathbf{d}}_{k}}={{\left[ {{d}_{k,1}},{{d}_{k,2}},\ldots ,{{d}_{k,L}} \right]}^{T}}\in {\mathbb{R}^{L\times \text{1}}}$ and a sensing symbol vector ${{\mathbf{s}}_{k}^{'}}={{\left[ {{s}_{k,1}^{'}},{{s}_{k,2}^{'}},\ldots ,{{s}_{k,J}^{'}} \right]}^{T}}\in {\mathbb{R}^{J\times 1}},k=1,...,K$ in each time slot, where the computation symbol ${{d}_{k,l}}$ and the sensing symbol ${s}_{k,j}^{'}$ are the measured values of the parameter $l$ and the parameter $j$ at the $k$th UE, respectively. For computation, the BS engages itself in computing $L$ nomographic functions \cite{Nomofun}, such that
\begin{equation}\label{eqnomo}
  q_l={{f}_{l}}\left( \sum\limits_{k=1}^{K}{{{g}_{k,l}}\left( {{d}_{k,l}} \right)} \right), l=1,...,L,
\end{equation}
where $f_l(\cdot)$ and $g_{k,l}(\cdot)$ represent post-processing functions at the BS and pre-processing functions at the IoT UEs, respectively (see Table \ref{Functions} for examples).
Let ${{\mathbf{s}}_{k}}={{\left[ {{g}_{k,1}}\left( {{d}_{k,1}} \right),{{g}_{k,2}}\left( {{d}_{k,2}} \right),\ldots ,{{g}_{k,L}}\left( {{d}_{k,L}} \right) \right]}^{T}}$ denote the transmitted computation signal after pre-processing at the $k$th IoT UE.  Thus, the $k$th UE constructs the superposition coded transmit signal ${{\mathbf{x}}_{k}}$ as
\begin{equation}\label{signal1}
  {{\mathbf{x}}_{k}}={{\mathbf{W}}_{k}}{{\mathbf{s}}_{k}}+\sum\limits_{j=1}^{J}{{{\mathbf{v}}_{k,j}}s_{k,j}^{'}},
\end{equation}
 where $\mathbf{W}_k\in \mathbb{C}^{M\times L}$ denotes the transmit computation beamforming matrix for the computation signal ${{\mathbf{s}}_{k}}$, and ${{\mathbf{v}}_{k,j}}\in \mathbb{C}^{M\times 1}$ is the transmit sensing beamforming vector for the sensing signal $s_{k,j}^{'}$. For ease of analysis but without loss of generality, we assume that $\mathrm{E}\left\{ {{\mathbf{s}}_{k}}\mathbf{s}_{k}^{H} \right\}=\mathbf{I}$ and $\mathrm{E}\left\{ {{s}_{k,j}^{'}}{{s}_{k,j}^{'H}} \right\}=1$. Therefore, the received signal at the BS is given by
 \begin{eqnarray}\label{receivesignal}
\mathbf{y}&=&\sum\limits_{k=1}^{K}{{{\mathbf{H}}_{k}}{{\mathbf{x}}_{k}}+\mathbf{n}}\nonumber\\
&=&\underbrace{\sum\limits_{k=1}^{K}{{{\mathbf{H}}_{k}}{{\mathbf{W}}_{k}}{{\mathbf{s}}_{k}}}}_{\text{computation signal}}+\underbrace{\sum\limits_{k=1}^{K}{\sum\limits_{j=1}^{J}{{{\mathbf{H}}_{k}}{{\mathbf{v}}_{k,j}}s_{k,j}^{'}}}}_{\text{sensing signal}}+\mathbf{n},
\end{eqnarray}
 where $\mathbf{n}$ is the additive white Gaussian noise (AWGN) vector with the distribution $\mathcal{CN}(\mathbf{0},\sigma_n^2\mathbf{I})$, and ${{\mathbf{H}}_{k}}\in \mathbb{C}^{N\times M}$ denotes the MIMO channel matrix from the $k$th UE to the BS. It is reasonably assumed that ${{\mathbf{H}}_{k}}$ remains unchanged during a time slot, but independently fades over time slots.

Firstly, we discuss the processing of the computation signal. Due to the one-to-one mapping between $\mathbf{s}=\sum\limits_{k=1}^{K}{{{\mathbf{s}}_{k}}}$ and $\mathbf{q}=[q_1,q_2,...,q_L]^T$ in (\ref{eqnomo}), we take an accurate $\mathbf{s}$ at the BS as the targeted function signal. To minimize the distortion of the targeted function signal caused by channel fading, noise and interference, it is necessary to perform receive beamforming at the BS. Thus, the received signal for computation at the BS is given by
\begin{equation}\label{signal2}
    \mathbf{\hat{s}}={{\mathbf{Z }}^{H}}\sum\limits_{k=1}^{K}{{{\mathbf{H}}_{k}}{{\mathbf{W}}_{k}}}{{\mathbf{s}}_{k}}+{{\mathbf{Z}}^{H}}\left( \sum\limits_{k=1}^{K}{{{\mathbf{H}}_{k}}\sum\limits_{j=1}^{J}{{{\mathbf{v}}_{k,j}}s_{k,j}^{'}}}+\mathbf{n} \right),
\end{equation}
where $\mathbf{Z}\in \mathbb{C}^{N\times L}$ denotes the receive computation beamforming matrix at the BS. Mathematically, the accuracy of computation at the BS can be measured by the MSE between $\mathbf{s}$ and $\mathbf{\hat{s}}$, which is given by
\begin{equation}\label{MSE1}
  \text{MSE}\left( \mathbf{\hat{s}},\mathbf{s} \right)\text{=}\mathrm{E}\left\{\mathrm{tr}\left( \left( \mathbf{\hat{s}}-\mathbf{s} \right){{\left( \mathbf{\hat{s}}-\mathbf{s} \right)}^{H}} \right) \right\}.
\end{equation}
Substituting (\ref{signal2}) into (\ref{MSE1}), the computation distortion can be expressed as the following MSE function in terms of receive and transmit beams:
\begin{eqnarray}\label{MSE2}
   \text{MSE}\left( \mathbf{Z},\mathbf{W}_k,\mathbf{v}_{k,j} \right)&=&\sum\limits_{k=1}^{K}{\left\| {{\mathbf{Z}}^{H}}{{\mathbf{H}}_{k}}{{\mathbf{W}}_{k}}-\mathbf{I} \right\|_{F}^{2}}+\sigma _{n}^{2}\left\| \mathbf{Z} \right\|_{F}^{2}\nonumber\\
   &+&\sum\limits_{k=1}^{K}{\sum\limits_{j=1}^{J}{{{\left\| {{\mathbf{Z}}^{H}}{{\mathbf{H}}_{k}}{{\mathbf{v}}_{k,j}} \right\|}^{2}}}}.
\end{eqnarray}
Secondly, we consider the processing of the sensing signal. The received signal of the $j$th sensing symbol sent from the $k$th UE at the BS is given by
\begin{eqnarray}\label{yID}
  y_{k,j}^{'}&=&\mathbf{u}_{k,j}^{H}{{\mathbf{H}}_{k}}{{\mathbf{v}}_{k,j}}s_{k,j}^{'}+\mathbf{u}_{k,j}^{H}\sum\limits_{i=1,i\ne k}^{K}{{{\mathbf{H}}_{i}}\sum\limits_{m=1,m\ne j}^{J}{{{\mathbf{v}}_{i,m}}s_{i,m}^{'}}}\nonumber\\
  &+&\mathbf{u}_{k,j}^{H}\sum\limits_{i=1}^{K}{{{\mathbf{H}}_{i}}}{{\mathbf{W}}_{i}}{{\mathbf{s}}_{i}}+\mathbf{u}_{k,j}^{H}\mathbf{n},
\end{eqnarray}
where $\mathbf{u}_{k,j} \in \mathbb{C}^{N\times 1}$ denotes the receive sensing beamforming vector of the $j$th sensing symbol for the $k$th UE at the BS. As a result, the corresponding received signal-to-interference-plus-noise ratio (SINR) at the sensing receiver can be expressed as (\ref{SINR}) at the top of next page.
\begin{figure*}
\begin{equation}\label{SINR}
{{\Gamma }_{k,j}}=\frac{{{\left| \mathbf{u}_{k,j}^{H}{{\mathbf{H}}_{k}}{{\mathbf{v}}_{k,j}} \right|}^{2}}}{\sum\limits_{i=1,i\ne k}^{K}{\sum\limits_{m=1,m\ne j}^{J}{{{\left| \mathbf{u}_{k,j}^{H}{{\mathbf{H}}_{i}}{{\mathbf{v}}_{i,m}} \right|}^{2}}}+\sum\limits_{i=1}^{K}{{{\left\| \mathbf{u}_{k,j}^{H}{{\mathbf{H}}_{i}}{{\mathbf{W}}_{i}} \right\|}^{2}}+\sigma _{n}^{2}{{\left\| \mathbf{u}_{k,j} \right\|}^{2}}}}}.
\end{equation}
\hrulefill
\end{figure*}
The SINR determines the capacity of the communication channel, and hence influences the accuracy of sensed information at the BS.

As seen from (\ref{MSE2}) and (\ref{SINR}), the overall performance is jointly affected by the transmit beams $\mathbf{W}_k$ and $\mathbf{v}_{k,j}$ at the IoT UEs, and receive beams $\mathbf{Z}$ and $\mathbf{u}_{k,j}$ at the BS. Thus, it makes sense to jointly design transmit and receive beamforming to improve the performance of both computation and sensing for B5G cellular IoT.

\section{Design of B5G Cellular IoT Integrating SCC}
In this section, we aim to jointly design transmit and receive beamforming matrices for B5G cellular IoT networks integrating SCC. Considering that IoT applications have different priorities between computation and sensing, we optimize the communication parameters from the perspectives of the computation error minimization and the weighted sum-rate maximization, respectively.

\subsection{Computation Error Minimization Design}
The design with the goal of minimizing the computation error of the computation signals in the B5G cellular IoT integrating SCC while guaranteeing the rate requirements of the sensing signals can be formulated as the following optimization problem:
\begin{subequations}\label{OP1}
\begin{eqnarray}
\underset{{{\mathbf{W}}_{k}},{{\mathbf{v}}_{k,j}},\mathbf{Z},{{\mathbf{u}}_{k,j}},\forall k,j}{\mathop{\min }}\,\!\!\!\!&&\!\!\!\! \text{MSE}\left( \mathbf{Z},\mathbf{W}_k,\mathbf{v}_{k,j} \right) \label{OP1obj}\\
\textrm{s.t.}\!\!\!\!&&\!\!\!\!\!{{\log }_{2}}\left( 1+{\Gamma }_{k,j} \right)\ge {{r}_{k,j}}, \label{OP1st1}\\
\!\!\!\!\!&&\!\!\!\!\!\left\| {{\mathbf{W}}_{k}} \right\|_{F}^{2}+\sum\limits_{j=1}^{J}{{{\left\| {{\mathbf{v}}_{k,j}} \right\|}^{2}}}\le {{P}_{\max ,k}},\label{OP1st2}
\end{eqnarray}
\end{subequations}
where $r_{k,j}$ is the required minimum achievable rate (in b/s) of the $j$th sensing signal at the $k$th UE, and $P_{\max,k}$ is the maximum transmit power budget at the $k$th UE. It is seen that the problem (\ref{OP1}) is NP-hard \cite{NP1,NP2}, and non-convex due to the coupled variables of transmit beams $\{\mathbf{W}_k,\mathbf{v}_k\}$ and receive beams $\{\mathbf{Z},\mathbf{u}_k\}$ in the objective function and the constraints. To solve this problem, we adopt an alternative optimization (AO) method to divide it into two subproblems, i.e., optimizing transmit beams with fixed receive beams, and optimizing receive beams with fixed transmit beams. The AO method stops until the value of the objective function for the original problem approaches a stationary point in the iterations.  Now, we first consider the subproblem for the optimization of receive beams. To balance the system performance and the design complexity, we employ minimum mean square error (MMSE) receivers,  which are given by
\begin{equation}\label{receiverZ}
  \mathbf{Z}={{\left( \sigma _{n}^{2}\mathbf{I}+\sum\limits_{k=1}^{K}{{{\mathbf{H}}_{k}}\Xi_k\mathbf{H}_{k}^{H}} \right)}^{-1}}\sum\limits_{k=1}^{K}{{{\mathbf{H}}_{k}}{{\mathbf{W}}_{k}}},
\end{equation}
and
\begin{equation}\label{receiverU}
{{\mathbf{u}}_{k,j}}={{\left( \sigma _{n}^{2}\mathbf{I}+\sum\limits_{k=1}^{K}{{{\mathbf{H}}_{k}}\Xi_k\mathbf{H}_{k}^{H}} \right)}^{-1}}{{\mathbf{H}}_{k}}{{\mathbf{v}}_{k,j}},
\end{equation}
respectively, where $\Xi_k={{\mathbf{W}}_{k}}\mathbf{W}_{k}^{H}+\sum\limits_{j=1}^{J}{{{\mathbf{v}}_{k,j}}\mathbf{v}_{k,j}^{H}}$.  Next, we deal with the other subproblem for the optimization of transmit beams $\mathbf{W}_k$ and $\mathbf{v}_{k,j}, \forall k,j$, with fixed receivers $\mathbf{Z}$ and $\mathbf{u}_{k,j}, \forall k,j$, in (\ref{receiverZ}) and (\ref{receiverU}),  which can be expressed as
\begin{eqnarray}
\!\!\!\!\!\!\!\!\underset{\begin{smallmatrix}
 {{\mathbf{W}}_{k}},{{\mathbf{v}}_{k,j}}, \\
 \forall k,j
\end{smallmatrix}}{\mathop{\min }}\!\!\!\!\,\!\!\!\!&&\!\!\!\!\sum\limits_{k=1}^{K}{\left\| {{\mathbf{Z }}^{H}}{{\mathbf{H}}_{k}}{{\mathbf{W}}_{k}}-\mathbf{I} \right\|_{F}^{2}}+\sum\limits_{k=1}^{K}{\sum\limits_{j=1}^{J}{{{\left\| {{\mathbf{Z }}^{H}}{{\mathbf{H}}_{k}}{{\mathbf{v}}_{k,j}} \right\|}^{2}}}} \label{OP2}\\
\textrm{s.t.}&&\!\!\!\!\!(\ref{OP1st1}),(\ref{OP1st2}).\nonumber
\end{eqnarray}
As seen from the problem (\ref{OP2}), the objective function and the transmit power constraint (\ref{OP1st2}) are convex, but the rate constraint (\ref{OP1st1}) is non-convex. To this end, we define ${{\gamma }_{k,j}}={{2}^{{{r}_{k,j}}}}-1$, and shift the terms to obtain
\begin{eqnarray}\label{SINR0}
\!\!\! \!\!\!\frac{1}{{{\gamma }_{k,j}}}{{\left| \mathbf{u}_{k,j}^{H}{{\mathbf{H}}_{k}}{{\mathbf{v}}_{k,j}} \right|}^{2}} \!\!\!&\ge&\!\!\! \sum\limits_{i=1,i\ne k}^{K}{\sum\limits_{m=1,m\ne j}^{J}{{{\left| \mathbf{u}_{k,j}^{H}{{\mathbf{H}}_{i}}{{\mathbf{v}}_{i,m}} \right|}^{2}}}}\nonumber\\
 \!\!\!&+&\!\!\!\sum\limits_{i=1}^{K}{{{\left\| \mathbf{u}_{k,j}^{H}{{\mathbf{H}}_{i}}{{\mathbf{W}}_{i}} \right\|}^{2}}}+\sigma _{n}^{2}{{\left\| {{\mathbf{u}}_{k,j}} \right\|}^{2}}.
\end{eqnarray}
To further address the non-convexity of the constraint (\ref{SINR0}), we introduce ${{\mathbf{V}}_{k,j}}={{\mathbf{v}}_{k,j}}\mathbf{v}_{k,j}^{H}$, and transform it as
\begin{eqnarray}\label{SINR1}
   \!\!\!\!\!\!\!\!\!&&\!\!\!\frac{1}{{{\gamma }_{k,j}}}\left( \mathbf{u}_{k,j}^{H}{{\mathbf{H}}_{k}}{{\mathbf{V}}_{k,j}}\mathbf{H}_{k}^{H}{{\mathbf{u}}_{k,j}} \right)\ge\sum\limits_{i=1}^{K}{{{\left\| \mathbf{u}_{k,j}^{H}{{\mathbf{H}}_{i}}{{\mathbf{W}}_{i}} \right\|}^{2}}}+\nonumber\\
   \!\!\!\!\!\!\!\!\!&&\!\!\!\sum\limits_{i=1,i\ne k}^{K}{\sum\limits_{m=1,m\ne j}^{J}\left({\mathbf{u}_{k,j}^{H}{{\mathbf{H}}_{i}}{{\mathbf{V}}_{i,m}}\mathbf{H}_{i}^{H}{{\mathbf{u}}_{k,j}}} \right)}+
  \sigma _{n}^{2}{{\left\| {{\mathbf{u}}_{k,j}} \right\|}^{2}}.
\end{eqnarray}
Accordingly, the problem (\ref{OP2}) can be reformulated as a semi-definite programming (SDP) problem (\ref{OP3}) at the top of next page.
\begin{figure*}
\begin{subequations}\label{OP3}
\begin{eqnarray}
\underset{{{\mathbf{W}}_{k}},{{\mathbf{v}}_{k,j}},\forall k,j}{\mathop{\min }}\,\!\!\!\!&&\!\!\!\!\sum\limits_{k=1}^{K}{\left\| {{\mathbf{Z }}^{H}}{{\mathbf{H}}_{k}}{{\mathbf{W}}_{k}}-\mathbf{I} \right\|_{F}^{2}}+\sum\limits_{k=1}^{K}{\sum\limits_{j=1}^{J}{\text{tr}\left( {{\mathbf{Z }}^{H}}{{\mathbf{H}}_{k}}{{\mathbf{V}}_{k,j}}\mathbf{H}_{k}^{H}\mathbf{Z } \right)}}\label{OP3obj}\\
\textrm{s.t.}&&\!\!\!\! (\ref{SINR1}),\nonumber\\
&&\!\!\!\!\left\| {{\mathbf{W}}_{k}} \right\|_{F}^{2}+\sum\limits_{j=1}^{J}{\text{tr}\left( {{\mathbf{V}}_{k,j}} \right)}\le {{P}_{\max ,k}},\label{OP3st1}\\
&&\!\!\!\!{{\mathbf{V}}_{k,j}}\succeq \mathbf{0},\forall k,j,\label{OP3st2}\\
&&\!\!\!\!\text{Rank}\left( {{\mathbf{V}}_{k,j}} \right)=1, \forall k,j.\label{OP3st3}
\end{eqnarray}
\end{subequations}
\hrulefill
\end{figure*}
However, the problem (\ref{OP3}) is still non-convex due to the rank-one constraint of ${{\mathbf{V}}_{k,j}}$. To address this issue, we apply the semi-definite relaxation (SDR) technique to this problem, i.e., dropping the rank-one constraint (\ref{OP3st3}). Therefore, the problem (\ref{OP3}) is reduced as
\begin{eqnarray}\label{OP3.5}
\underset{{{\mathbf{W}}_{k}},{{\mathbf{V}}_{k,j}},\forall k,j}{\mathop{\min }}\,\!\!\!\!&&\!\!\!\!\sum\limits_{k=1}^{K}{\left\| {{\mathbf{Z }}^{H}}{{\mathbf{H}}_{k}}{{\mathbf{W}}_{k}}-\mathbf{I}  \right\|_{F}^{2}}\nonumber\\
&+&\sum\limits_{k=1}^{K}{\sum\limits_{j=1}^{J}{\text{tr}\left( {{\mathbf{Z }}^{H}}{{\mathbf{H}}_{k}}{{\mathbf{V}}_{k,j}}\mathbf{H}_{k}^{H}\mathbf{Z } \right)}}\\
\textrm{s.t.}&&\!\!\!\! (\ref{SINR1}),(\ref{OP3st1}),(\ref{OP3st2}).\nonumber
\end{eqnarray}
In terms of transmit beams $\{\mathbf{W}_k,\mathbf{v}_k\}$, the objective function in (\ref{OP3.5}) is a convex function and the constraints (\ref{SINR1}), (\ref{OP3st1}) and (\ref{OP3st2}) are all convex sets. Thus, the problem (16) is convex, which can be effectively solved by some optimization tools, such as CVX \cite{CVX}. It is worth mentioning that dropping the rank-one constraint (\ref{OP3st3}) does not affect the optimal solution to the problem (\ref{OP3.5}). Specifically, for the optimal solution ${{\mathbf{V}}_{k,j}^{*}}$ to the problem (\ref{OP3.5}), we have the following theorem.

\emph{Theorem 1:} The optimal solution ${{\mathbf{V}}_{k,j}^{*}}$ of the problem (\ref{OP3.5}) always satisfies $\text{Rank}\left( {{\mathbf{V}}_{k,j}^{*}} \right)=1, \forall k,j$.
\begin{IEEEproof}
Please refer to Appendix A.
\end{IEEEproof}

Hence, we can recover the optimal transmit sensing beams ${{\mathbf{v}}_{k,j}^{*}},\forall k,j$ of the original problem (\ref{OP1}) via eigenvalue decomposition (EVD) on ${{\mathbf{V}}_{k,j}^{*}}$, namely
\begin{equation}\label{EVD}
  {{\mathbf{v}}_{k,j}^{*}}=\sqrt{{{\lambda }_{k,j}^{\max }}\left( {{\mathbf{V}}_{k,j}^{*}} \right)}{{\bm{\xi }}_{k,j}^{\max }},
\end{equation}
where ${{\lambda }_{k,j}^{\max }}\left( {{\mathbf{V}}_{k,j}^{*}} \right)$ is the maximum eigenvalue of $\mathbf{V}_{k,j}^*$ and ${{\bm{\xi }}_{k,j}^{\max }}$ is the unit eigenvector related to ${{\lambda }_{k,j}^{\max }}\left( {{\mathbf{V}}_{k,j}^{*}} \right)$.  In summary, the design of B5G cellular IoT integrating SCC for minimizing the computation error can be described as Algorithm 1.

\begin{algorithm}[h]
\caption{: B5G Cellular IoT Integrating SCC Design for Computation Error Minimization.}
\label{alg1}
\hspace*{0.02in} {\bf Input:} 
 $N,K,M,L,J,\sigma_n^2,r_{k,j},P_{\max,k},\forall k,j$\\
\hspace*{0.02in} {\bf Output:} 
$\mathbf{W}_k, \mathbf{v}_{k,j},\mathbf{Z},\mathbf{u}_{k,j},\forall k,j$

\begin{algorithmic}[1]
\STATE{\textbf{Initialize} $\mathbf{W}_k^{(0)}=[\sqrt{\frac{P_{\max,k}}{2}},0,\ldots,0]^{T}\times[1,0,\ldots,0]$, $\mathbf{v}_{k,j}^{(0)}=[\sqrt{\frac{P_{\max,k}}{2J}},0,\ldots,0]^{T},\forall k,j$, iteration index $t=0$;}
\REPEAT
 \STATE{compute $\mathbf{Z}^{(t)}$ by (\ref{receiverZ}) with $\mathbf{W}_k^{(t-1)}$ and $\mathbf{v}_{k,j}^{(t-1)}$;}
 \STATE{compute $\mathbf{u}_{k,j}^{(t)}$ by (\ref{receiverU}) with $\mathbf{W}_k^{(t-1)}$ and $\mathbf{v}_{k,j}^{(t-1)}$;}
 \STATE{obtain $\{\mathbf{W}^{(t)}_{k},\mathbf{V}_{k,j}^{(t)}\}$ by solving the problem (\ref{OP3.5}) via CVX with fixed $\{\mathbf{Z}^{(t)},\mathbf{u}_{k,j}^{(t)}\}$; }
 \STATE{obtain  $\mathbf{v}_{k,j}^{(t)}$ by EVD on $\mathbf{V}_{k,j}^{(t)}$ according to (\ref{EVD});}
 \STATE{$t=t+1$;}
\UNTIL{convergence}
 \end{algorithmic}
\end{algorithm}

\subsection{Weighted Sum-rate Maximization Design}
In this section, we design the B5G cellular IoT integrating SCC from the perspective of maximizing the weighted sum-rate of sensing signals, while fulfilling the requirement of the computation error of computation signals. The design is formulated as the following optimization problem:
\begin{subequations}\label{OP4}
\begin{eqnarray}
\underset{{{\mathbf{W}}_{k}},{{\mathbf{v}}_{k,j}},\mathbf{Z},{{\mathbf{u}}_{k,j}},\forall k,j}{\mathop{\max }}\,\!\!\!\!&&\!\!\!\!\sum\limits_{k=1}^{K}{\sum\limits_{j=1}^{J}{{{\theta }_{k,j}}{{\log }_{2}}\left( 1+{{\Gamma }_{k,j}} \right)}}\label{OP3obj}\\
\textrm{s.t.}&&\!\!\!\! (\ref{OP1st2}),\nonumber\\
&&\!\!\!\!\text{MSE}\left( \mathbf{Z},\mathbf{W}_k,\mathbf{v}_{k,j} \right)\le \rho ,\label{OP4st1}
\end{eqnarray}
\end{subequations}
where ${\theta }_{k,j}>0$ denotes the priority of the sensing signal $s_{k,j}^{'}$, and $\rho$ is the maximum tolerable computational error. Similarly, the formulated problem is NP-hard \cite{NP3,NP4}, and nonconvex due to the coupled transmit beams and receive beams, for which finding its optimal solution in polynomial time is difficult. Hence, the problem is split into two subproblems by using the AO method. For the subproblem of optimizing receive beams, we also adopt the MMSE receivers in (\ref{receiverZ}) and (\ref{receiverU}). Thus, the other subproblem of optimizing transmit beams can be expressed as
\begin{eqnarray}\label{OP5}
\underset{{{\mathbf{W}}_{k}},{{\mathbf{v}}_{k,j}},\forall k,j}{\mathop{\max }}\,\!\!\!\!&&\!\!\!\!\sum\limits_{k=1}^{K}{\sum\limits_{j=1}^{J}{{{\theta }_{k,j}}{{\log }_{2}}\left( 1+{{\Gamma }_{k,j}} \right)}}\\
\textrm{s.t.}&&\!\!\!\! (\ref{OP1st2}),(\ref{OP4st1}).\nonumber
\end{eqnarray}
It is seen from (\ref{OP5}) that the weighted sum-rate maximization in terms of optimizing transmit beams is a non-convex problem due to the complicated non-convex objective function. To solve this issue, we handle its objective function by applying the following theorem:

\emph{Theorem 2:} The received SINR $\Gamma_{k,j}$ and the MMSE $e_{k,j}$ for the sensing signal $s_{k,j}^{'}$ have the relationship of $1+{{\Gamma }_{k,j}}=e_{k,j}^{-1}$.
\begin{IEEEproof}
Please refer to Appendix B.
\end{IEEEproof}
Based on Theorem 2, the objective function of the problem (\ref{OP5}) is changed as
\begin{equation}\label{obj1}
  \underset{{{\mathbf{W}}_{k}},{{\mathbf{v}}_{k,j}},\forall k,j}{\mathop{\min }}\,\sum\limits_{k=1}^{K}{\sum\limits_{j=1}^{J}{{{\theta }_{k,j}}{{\log }_{2}}\left( e_{k,j} \right)}}.
\end{equation}
Note that the transformed objective function (\ref{obj1}) is still nonconvex, but is equivalent to minimizing a function of MSE with the MMSE receiver. Actually, we use exactly the MMSE receivers $\mathbf{u}_{k,j}, \forall k,j$ in the subproblem of optimizing receive beams. Thus, (\ref{obj1}) can be reformulated as
 \begin{equation}\label{obj2}
  \underset{{{\mathbf{W}}_{k}},{{\mathbf{v}}_{k,j}},\forall k,j}{\mathop{\min }}\,\sum\limits_{k=1}^{K}{\sum\limits_{j=1}^{J}{{{\theta }_{k,j}}{{\log }_{2}}\left( \mathrm{MSE}_{k,j} \right)}},
\end{equation}
where $\mathrm{MSE}_{k,j}$ is the MSE related to the sensing signal $s_{k,j}^{'}$, which is given by
\begin{eqnarray}\label{MSEkj}
  \mathrm{MSE}_{k,j}\!\!\!&=&\!\!\!\mathbf{u}_{k,j}^{H}\left( \sum\limits_{i=1}^{K}{\sum\limits_{m=1}^{J}{{{\mathbf{H}}_{i}}{{\mathbf{v}}_{i,m}}\mathbf{v}_{i,m}^{H}}\mathbf{H}_{i}^{H}} \right){{\mathbf{u}}_{k,j}}\nonumber\\
  \!\!\!&+&\!\!\!\mathbf{u}_{k,j}^{H}\left( \sum\limits_{i=1}^{K}{{{\mathbf{H}}_{i}}{{\mathbf{W}}_{i}}\mathbf{W}_{i}^{H}\mathbf{H}_{i}^{H}}+\sigma _{n}^{2}\mathbf{I} \right){{\mathbf{u}}_{k,j}}\nonumber\\
   \!\!\!&-&\!\!\!\mathbf{u}_{k,j}^{H}{{\mathbf{H}}_{k}}{{\mathbf{v}}_{k,j}}-\mathbf{v}_{k,j}^{H}\mathbf{H}_{k}^{H}{{\mathbf{u}}_{k,j}}+1.
\end{eqnarray}
The detailed derivation of $\mathrm{MSE}_{k,j}$ can be found in Appendix B. However, the sum of logarithmic function (\ref{obj2}) keeps us from further addressing its nonconvexity. To this end, we introduce a weight variable $\beta_{k,j}$ for the MMSE receiver $\mathbf{u}_{k,j}$ \cite{WMMSE,WMMSE2}, and thus the logarithmic function can be replaced by the following term:
 \begin{equation}\label{obj3}
  \underset{{{\mathbf{W}}_{k}},{{\mathbf{v}}_{k,j}},\beta_{k,j},\forall k,j}{\mathop{\min }}\,\sum\limits_{k=1}^{K}{\sum\limits_{j=1}^{J}{{{\theta }_{k,j}}\left( \beta_{k,j}\mathrm{MSE}_{k,j}-{{\log }_{2}}(\beta_{k,j}) \right)}}.
\end{equation}
Note that (\ref{obj2}) and (\ref{obj3}) are equivalent when $\beta_{k,j}=\mathrm{MSE}_{k,j}^{-1}$. Combined with the subproblem of optimizing receive beams, the original optimization problem (\ref{OP4}) can eventually be transformed as
\begin{eqnarray}\label{OP6}
\!\!\!\!\underset{\begin{smallmatrix}
 \mathbf{Z},{{\mathbf{u}}_{k,j}},{{\beta }_{k,j,}} \\
 {{\mathbf{W}}_{k}},{{\mathbf{v}}_{k,j}},\forall k,j
\end{smallmatrix}}{\mathop{\min }}\!\!\!\,\!\!\!\!&&\!\!\!\!\sum\limits_{k=1}^{K}{\sum\limits_{j=1}^{J}{{{\theta }_{k,j}}\left( {{\beta }_{k,j}}\mathrm{MSE}_{k,j}-{{\log }_{2}}\left( {{\beta }_{k,j}} \right) \right)}}\\
\textrm{s.t.}&&\!\!\!\! (\ref{OP1st2}),(\ref{OP4st1}).\nonumber
\end{eqnarray}
 It is found that the problem (\ref{OP6}) is not a joint convex function of $\mathbf{Z},{{\mathbf{u}}_{k,j}},{{\mathbf{W}}_{k}},{{\mathbf{v}}_{k,j}},\beta_{k,j},\forall k,j$, but is a convex function in each of $\{{{\mathbf{Z}}},{{\mathbf{u}}_{k,j}}\}$, $\{{{\mathbf{W}}_{k}},{{\mathbf{v}}_{k,j}}\}$ and $\beta_{k,j}$, respectively. Hence, we can adopt the block coordinate decent method to solve this problem. Specifically, we sequentially optimize one variable by fixing the others until they approach a stationary point in the iterations. First, there are closed-form solutions for the MMSE receivers $\{{{\mathbf{Z}}},{{\mathbf{u}}_{k,j}}\}$ according to (\ref{receiverZ}) and (\ref{receiverU}). Second, for the weight variable $\beta_{k,j}$, we set $\beta_{k,j}=\mathrm{MSE}_{k,j}^{-1}$ with the MMSE receivers. Finally, since the transmit beams $\{{{\mathbf{W}}_{k}},{{\mathbf{v}}_{k,j}}\}$ involve multiple convex constraints, i.e., (\ref{OP1st2}) and (\ref{OP4st1}), they can be directly solved by CVX. In summary, the design of B5G cellular IoT integrating SCC for maximizing the weighted sum-rate can be described as Algorithm 2.
\begin{algorithm}[h]
\caption{: B5G Cellular IoT Integrating SCC Design for Weighted Sum-rate Maximization.}
\label{alg2}
\hspace*{0.02in} {\bf Input:} 
 $N,K,M,L,J,\sigma_n^2,\rho,P_{\max,k},\forall k$\\
\hspace*{0.02in} {\bf Output:}
$\mathbf{W}_k, \mathbf{v}_{k,j},\mathbf{Z},\mathbf{u}_{k,j},\forall k,j$

\begin{algorithmic}[1]
\STATE{\textbf{Initialize} $\mathbf{W}_k^{(0)}=[\sqrt{\frac{P_{\max,k}}{2}},0,\ldots,0]^{T}\times[1,0,\ldots,0]$, $\mathbf{v}_{k,j}^{(0)}=[\sqrt{\frac{P_{\max,k}}{2J}},0,\ldots,0]^{T},\forall k,j$, iteration index $t=0$;}
\REPEAT
 \STATE{compute $\mathbf{Z}^{(t)}$ by(\ref{receiverZ}) with $\mathbf{W}_k^{(t-1)}$ and $\mathbf{v}_{k,j}^{(t-1)}$;}
 \STATE{compute $\mathbf{u}_{k,j}^{(t)}$ by (\ref{receiverU}) with $\mathbf{W}_k^{(t-1)}$ and $\mathbf{v}_{k,j}^{(t-1)}$;}
 \STATE{update $\mathrm{MSE}_{k,j}^{(t)}$ by (\ref{MSEkj}), and set $\beta_{k,j}^{(t)}=1/\mathrm{MSE}_{k,j}^{(t)}$ ;}
 \STATE{obtain $\{\mathbf{W}^{(t)}_{k},\mathbf{v}_{k,j}^{(t)}\}$ by solving the problem (\ref{OP6}) via CVX with fixed $\{\mathbf{Z}^{(t)},\mathbf{u}_{k,j}^{(t)},\beta_{k,j}^{(t)}\}$; }
 \STATE{$t=t+1$;}
\UNTIL{convergence}
 \end{algorithmic}
\end{algorithm}

\subsection{Analysis of Algorithms}
In this section, we analyze the two proposed algorithms from the viewpoints of initialization analysis, convergence analysis and complexity analysis, respectively.
 \begin{enumerate}
\item \emph{Initialization Analysis:} Since the proposed algorithms are both iterative in nature, the initialization is important to achieve a quick convergence, and affects the final performance. First, for Algorithm 1, we adopt the MMSE receivers $\{\mathbf{Z},\mathbf{u}_{k,j}\}$, which are updated according to transmit beams $\{\mathbf{W}_k,\mathbf{v}_{k,j}\},\forall k,j$, in the iterations.
Since the optimization variables of the problem (\ref{OP3.5}) are subject to multiple constraints, the initialization of variables cannot be random. It is seen that the constraint (\ref{SINR1}) is jointly affected by transmit beams and receive beams. Thus, we set the initial value by focusing on the constraints (\ref{OP3st1}) and (\ref{OP3st2}) only related to the transmit beams. Without loss of generality, we make $\mathbf{W}_{k}^{(0)}=\sqrt{\frac{{{P}_{\max,k}}}{2}}\mathbf{x}_1\mathbf{x}_2$ and $\mathbf{v}_{k,j}^{(0)}=\sqrt{\frac{{{P}_{\max,k}}}{2J}}\mathbf{x}_1$ to satisfy the transmit power constraints  (\ref{OP3st1}) at the IoT UEs and semidefinite constraints (\ref{OP3st2}) on $\mathbf{V}_{k,j},\forall k,j$,
where $\mathbf{x}_1=[1,0,\ldots,0]^T \in {\mathbb{R}^{M\times \text{1}}}$ and $\mathbf{x}_2=[1,0,\ldots,0] \in {\mathbb{R}^{1\times \text{L}}}$. Similarly, since (\ref{OP4st1}) to the problem (\ref{OP6}) depends on the combined effects of transmit beams and receive beams, we set the same initial value of transmit beams to meet the transmit power constraint (\ref{OP1st2}) at the beginning of Algorithm 2.

 \item \emph{Convergence Analysis:} For Algorithm 1, it is seen that the algorithm works as long as the initial value is set properly. First, the adopted MMSE receivers are able to minimize the computation error. Then, since the problem (\ref{OP3.5}) is convex in terms of $\{\mathbf{W}_k,\mathbf{V}_{k,j}\},\forall k,j$, it is feasible to find the optimal solutions for minimizing the objective value via CVX directly. Thus, based on the steps in Algorithm 1, the solutions in the $t$-th iteration are feasible in the $(t+1)$-th iteration for the original problem (\ref{OP1}), which means that the objective value obtained in the $(t+1)$-th iteration is less than that in the $t$-th iteration. In other words, the computation error monotonically decreases after each iteration. Furthermore, due to the existence of the transmit power constraints (\ref{OP3st1}) at the IoT UEs, the computation error is lower bounded. According to the monotone bounded convergence theorem, Algorithm 1 is convergent. Analogously for Algorithm 2, since the problem (\ref{OP6}) in terms of each optimization variable is convex, the optimal solutions to each sub-problem can be obtained by CVX for each iteration, which indicates the solutions in the $t$-th iteration are the feasible solution in the $(t+1)$-th iteration. That guarantees the objective value obtained in the $(t+1)$-th iteration is greater than that in the $t$-th iteration, namely the weighted sum-rate monotonically increases after each iteration. Besides, owing to the constraints of transmit power at the IoT UEs, the weighted sum-rate has an upper bound. Hence, Algorithm 2 is also convergent. Based on the convergence rate analysis of AO in \cite{rate1} and \cite{rate2}, we can obtain that the convergence rates of the two proposed algorithms both show a two-stage behavior. Specifically, at first, the error decreases q-linearly until sufficiently small. After that, sub-linear convergence is initiated.  However, for the required number of iteration, it is a complex function of multiple system parameters. Generally, it is difficult to obtain the scaling law of the number of iterations. Thus, we show the required number of iterations under various SNRs by simulations in Fig. \ref{IterCE} and Fig. \ref{IterWSR}. It is seen that the proposed algorithms have fast convergence speed under different conditions.

\item \emph{Complexity Analysis:} By observing Algorithm 1 and Algorithm 2, it is known that the operation steps of each iteration are the same, and thus we only discuss the complexity of per-iteration for Algorithm 1 and Algorithm 2 in the following. For Algorithm 1, it is seen that the computational complexity depends largely on the step 5, namely obtaining  $\{\mathbf{W}_{k},\mathbf{V}_{k,j}\}$ by solving the SDP problem (\ref{OP3.5}). Similarly, the primary computational complexity of Algorithm 2 stems from the step 6, i.e., optimizing the transmit beams by solving the second-order cone program (SOCP) problem (\ref{OP6}) in terms of optimizing transmit beams. Since the problem (\ref{OP3.5}) and the problem (\ref{OP6}) both involve only linear matrix inequality (LMI) and second-order cone (SOC) constraints, they can be effectively solved through a standard interior-point method (IPM) \cite{Convex}. To be specific, the problem (\ref{OP3.5}) has $(KJ+K)$ LMI constraints of dimension $M$, and $KJ$ SOC constraints of dimension $M$. For the problem (\ref{OP6}), it has $K$ LMI constraints of dimension $M$, and $1$ SOC constraint of dimension $M$. Thus, for a given precision $\epsilon>0$ of solution, the worst-case per-iteration complexities of solving the problem (\ref{OP3.5}) and the problem (\ref{OP6}) by using a generic IPM are in order of $\ln(1/\epsilon)\varpi_1$ and $\ln(1/\epsilon)\varpi_2$, respectively \cite{complexity}. The detailed expressions of $\varpi_1$ and $\varpi_2$ are listed in Table \ref{Comp}. In addition, to visualize the complexity, we present the running time for per-iteration of Algorithm 1 and Algorithm 2 with different numbers of IoT UEs, which is solved by CVX at the SDPT3 solver in the Windows 7 operating system, cf. Table \ref{complexity} at the top of next page, where we set $N=64,K=32,M=2,L=1,J=1,\rho=0.3,r_{m,n}=r_0=0.5$ bit/s/Hz, and $P_{\max,k}=P_{0}=1$ W. 

    \begin{table*}[ht]
    \small
    \centering
    \caption{The worst-case per-iteration complexities of Algorithm 1 and Algorithm 2  }\label{Comp}
    \begin{tabular}{|c|c|}
    \hline
    Algorithms & Complexity is in order of $\ln \left( 1/\varepsilon  \right)\varpi_{\{1,2\}}$, where decision variable $n=\mathcal{O}(KM^2)$ ) \\ \hline
    Algorithm 1 & $\varpi_1=\sqrt{(KJM+2KJ+KM)}\cdot n \cdot\left[ \left( KJ+K \right)M^{2}(M+n)+{KJ{M}^{2}+n^2} \right]$  \\\hline
    Algorithm 2 & $\varpi_2=\sqrt{(KM+2)}\cdot n \cdot\left[KM^2(M+n)+M^2+n^2 \right]$  \\
    \hline
    \end{tabular}
    \end{table*}

    \begin{table*}[ht]
    \centering
    \caption{The running time (s) versus the number of IoT UEs for per-iteration of Algorithm 1 and Algorithm 2}\label{complexity}
    \begin{tabular}{|c|c|c|c|c|c|c|c|c|}
    \hline
    $K$ & 4 & 8 &12&16&20&24&28&32  \\\hline
    Algorithm 1 &4.2025 &6.1225 &12.2593& 20.3416 &31.1224 &44.5496 &60.6612&79.8184 \\\hline
    Algorithm 2 & 3.1211 &7.5169&14.2356&23.3528&34.8264&48.9062&65.5694&85.3184 \\
    \hline
    \end{tabular}
    \end{table*}
    \end{enumerate}

\section{Simulation Results}
\begin{table*}
\small
\centering
\caption{Simulation Parameters }\label{Simulation}
\begin{tabular}{|c|c|}
\hline
Parameters & Values \\ \hline
Number of BS antennas& $N=64$  \\\hline
IoT UEs & $K=32,M=2,L=1,J=1$ \\\hline
Cell radius    & $500 \text{ m}$ \\\hline
Priority of IoT UEs & $\theta_{m,n}=\theta=1$ \\\hline
Maximum transmit SNR at the IoT UEs & $\mathrm{SNR}=5$ dB \\\hline
Noise powers & $\sigma_n^2=-50$ dBm \\\hline
Minimum required rate threshold & $r_{m,n}=r_0=0.5$ bit/s/Hz \\\hline
Maximum tolerable computation error & $\rho=0.3$  \\\hline
\end{tabular}
\end{table*}

In this section, we provide extensive simulation results to validate the effectiveness of the proposed algorithms for B5G cellular IoT integrating SCC. Without loss of generality, it is assumed that all IoT UEs are randomly distributed within a range of the cell, and have the same maximum transmit power $P_{\max,k}=P_0$. To be close to the reality, the pass loss is modeled as $\mathrm{PL}_{\mathrm{dB}}=128.1+37.6\log_{10}(d)$ \cite{pathlossmodel}, where $d$ (km) is the distance between the BS and the IoT UE. For ease of analysis, we refer to normalized computation error $\mathrm{MSE}/K$ as the performance metric for computation, and use $\mathrm{SNR}=10\log_{10}(P_0/\sigma_n^2)$ to denote the transmit signal-to-noise ratio (SNR) (in dB).
Unless extra specification, the simulation parameters are set as in Table \ref{Simulation}.

\begin{figure}[h]
 \centering
\includegraphics [width=0.5\textwidth] {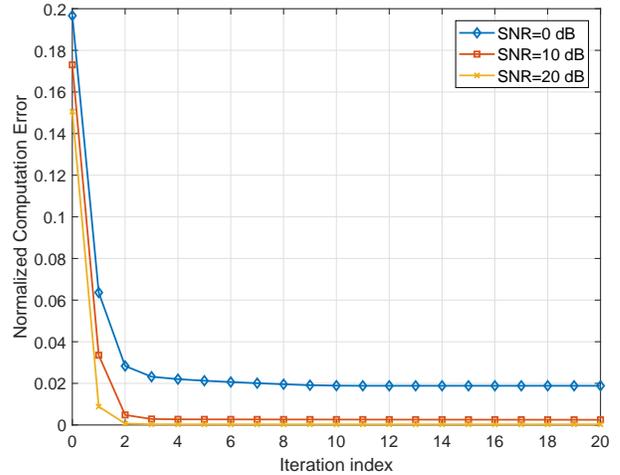}
\caption {Convergence behavior of the proposed Algorithm 1.}
\label{IterCE}
\end{figure}
First, we present the convergence behaviors of Algorithm 1 under different transmit SNR at the IoT UEs in Fig. \ref{IterCE}.
 It is seen that the computation error decreases monotonically as expected, and converges to its stationary value within few iterations on average under different transmit SNR at the IoT UEs. Moreover, Algorithm 1 has a steady convergence at high transmit SNR, while it requires more iterations at low transmit SNR region. This is because the received signal at the BS contains much more interference affecting the performance at low transmit SNR. In addition, the average running time of per-iteration for different number of IoT UEs is shown in Table \ref{complexity}. Combining the number of iterations and the running time for per-iteration, it is known that the complexity cost of Algorithm 1 is affordable for practical implementation.

\begin{figure}[h]
 \centering
\includegraphics [width=0.5\textwidth] {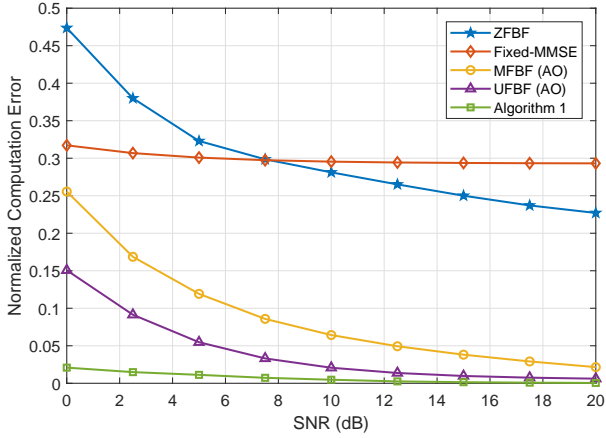}
\caption {Normalized computation error versus SNR (dB) for different algorithms.}
\label{dcompare64}
\end{figure}
Then, we show the performance gain of Algorithm 1 over four baseline beamforming algorithms in Fig. \ref{dcompare64}, i.e., a fixed MMSE algorithm with the fixed MMSE receivers only related to the channels between the BS and IoT UEs, a zero-forcing beamforming (ZFBF) with the zero-forcing transmitters, a match filtering beamforming (MFBF) based on the AO method with the match filtering receivers, and an uniform-forcing beamforming (UFBF) based on the AO method with the uniform-forcing transmitters and the MMSE receivers. As the SNR increases, the computation error decreases for all five algorithms. Since the fixed MMSE algorithm and the ZFBF algorithm are based on the fixed transmitters/receivers, they perform worse than the three AO algorithms. It is found that the fixed MMSE algorithm is superior to the ZFBF algorithm on the performance in the low SNR region, while performs worse than the ZFBF algorithm in the high SNR region. Moreover, it is seen that in the whole SNR region, although the UFBF algorithm outperforms the MFBF algorithm, the proposed Algorithm 1 can achieve the best performance. This is because it jointly optimizes the receive beamforming and transmit beamforming in the optimal way. That confirms the effectiveness of the proposed Algorithm 1.

\begin{figure}[h]
 \centering
\includegraphics [width=0.5\textwidth] {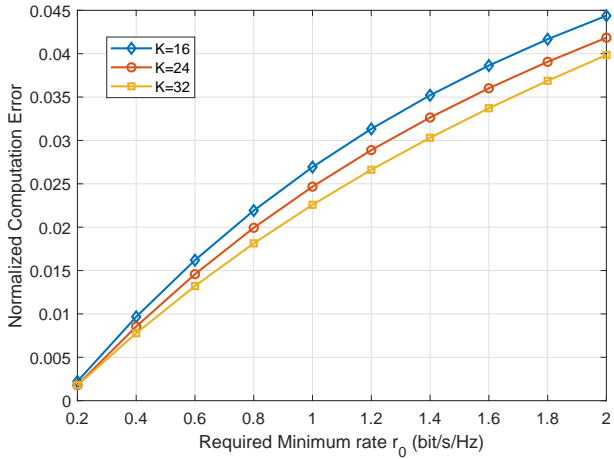}
\caption {Normalized computation error versus required minimum rate (bit/s/Hz) for different number of IoT UEs of the proposed Algorithm 1.}
\label{CE_rateK}
\end{figure}
Next, we check the impacts of the required minimum rate $r_0$ for each IoT UE and the number of IoT UEs $K$ on the computation error for Algorithm 1 in  Fig. \ref{CE_rateK}. It is found that for a given transmit SNR at IoT UEs, the normalized computation error increases with the increment of the required minimum rate. This is because the higher required minimum rate consumes more power to meet the sensing performance, resulting in less power to reduce the computation error. Thus, it is likely to enhance the computation performance with the limited transmit power by relaxing the sensing requirements. Moreover, as the number of IoT UEs $K$ increases, the normalized computation error decreases, since the combined received signal will be accordingly enhanced. Thus, Algorithm 1 can achieve more performance gains when the number of accessed IoT UEs is large, which exactly means Algorithm 1 is quite suitable to B5G cellular IoT with a massive number of devices.

\begin{figure}[h]
 \centering
\includegraphics [width=0.5\textwidth] {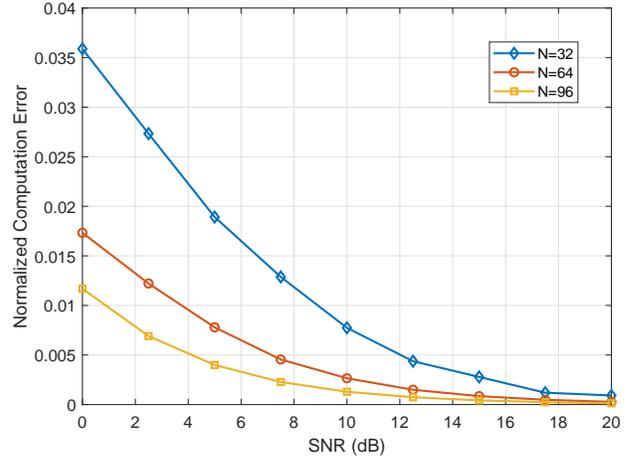}
\caption {Normalized computation error versus SNR (dB) for different number of antennas at the BS of the proposed Algorithm 1.}
\label{NT326496}
\end{figure}

Fig. \ref{NT326496} investigates the effect of the number of BS antennas $N$ on the computation error of Algorithm 1. For a given transmit SNR, the algorithm with more antennas at the BS can achieve a lower computation error due to more array gains. Moreover, the performance gains by adding more BS antennas decreases as the transmit SNR increases. Thus, it is able to enhance the performance for Algorithm 1 by increasing the BS antennas in the low and medium transmit SNR region. In addition, even with a not so large number of BS antennas, e.g., $N=32$, Algorithm 1 can obtain good performance by improving the transmit power at the IoT UEs.

\begin{figure}[h]
 \centering
\includegraphics [width=0.5\textwidth] {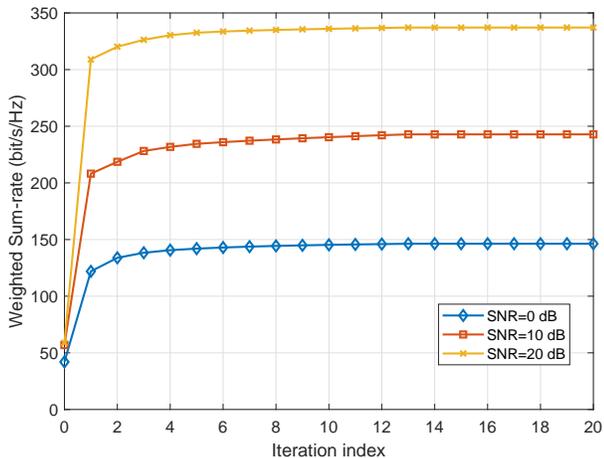}
\caption {Convergence behavior of the proposed Algorithm 2.}
\label{IterWSR}
\end{figure}
Fig. \ref{IterWSR} shows the convergence behaviors of Algorithm 2 under different SNR values. It is seen that the weighted sum-rate gradually improves as the number of iterations increases, and then stabilizes to an equilibrium point after no more than 10 iterations on average under different transmit SNR values, which means that the complexity of Algorithm 2 is affordable due to a low computational cost of per-iteration. In addition, for ease of observation, the average running time vs the number of IoT UEs on per-iteration of Algorithm 2 is presented in Table \ref{complexity}.

\begin{figure}[h]
 \centering
\includegraphics [width=0.5\textwidth] {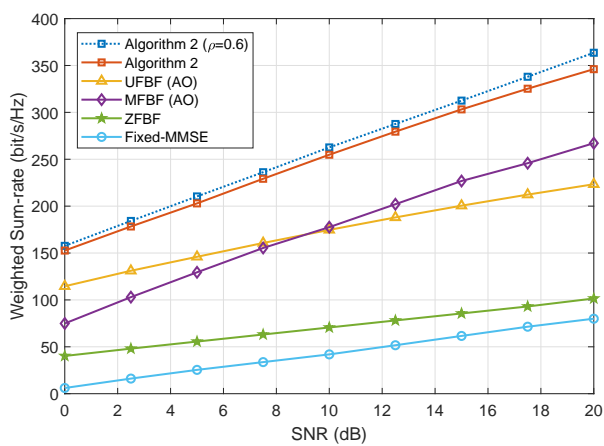}
\caption {Weighted sum-rate versus SNR (dB) for different algorithms.}
\label{WSRcomp}
\end{figure}

In Fig. \ref{WSRcomp}, we compare the weighted sum-rate versus SNR (dB) for different algorithms, i.e., a fixed MMSE algorithm, a ZFBF algorithm, a MFBF algorithm based on the AO method, an UFBF algorithm based on the AO method, and the proposed Algorithm 2. For all five algorithms, the weighted sum-rate improves as the SNR increases. It is seen that the two fixed algorithms (fixed MMSE algorithm and ZFBF algorithm) also exhibit poor performance compared to the three AO algorithms in the sense of maximizing the weighted sum-rate. Moreover, the proposed Algorithm 2 performs much better than the other two AO algorithms in the whole SNR region, which affirms the effectiveness of Algorithm 2. In addition, it is found that Algorithm 2 can achieve a higher weighted sum-rate with a high maximum tolerable computation error $\rho$. This is because for a given transmit power, a low requirement of the maximum tolerable computation error leads to more power consumed for the sensing performance, and less power for enhancing the computation performance. With the increasing of transmit power, the IoT UEs have enough power to meet the sensing requirements, and thus the performance gap between $\rho=0.6$ and $\rho=0.3$ enlarges as the increment of the SNR.

\begin{figure}[h]
 \centering
\includegraphics [width=0.5\textwidth] {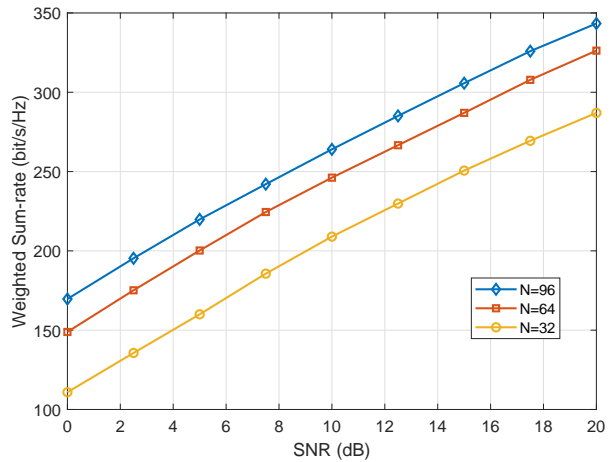}
\caption {Weighted sum-rate versus SNR (dB) for different BS antennas of the proposed Algorithm 2.}
\label{NtWSR}
\end{figure}
Finally, we check the influence of the number of BS antennas on the weighted sum-rate of Algorithm 2. As seen in Fig. \ref{NtWSR}, the weighted sum-rate improves by increasing the number of BS antennas, since more antennas at the BS can provide more array gains for performance enhancement. Moreover, it is found that the performance gap between $N=96$ and $N=64$ is smaller than that between $N=64$ and $N=32$, which means the performance gain by only adding BS antennas is limited. Besides, the BS equipped with a large-scale antenna array requires a huge cost in the practical system, e.g., radio frequency (RF) chain, although it does not cause higher computational complexity of Algorithm 2. Thus, it makes sense to select a suitable number of antennas at the BS according to system parameters and channel conditions of B5G cellular IoT integrating SCC.

\section{Conclusion}
This paper has designed a comprehensive framework for B5G cellular IoT integrating SCC. For realizing accurate computation and sensing with a massive of IoT UEs, two joint beamforming design algorithms for communications were proposed from the perspectives of minimizing the computation error while ensuring the rate requirements of sensing signals and maximizing the weighted sum-rate while guaranteeing the precision of computation results, respectively. Extensive simulations validated the effectiveness of the proposed algorithms for B5G cellular IoT integrating SCC.

\begin{appendices}
\section{The Proof of Theorem 1}
Before the proof, let us list two lemmas to be used.

\emph{Lemma 1 (Sylvester's rank inequality, \cite{rankieq})}: For matrix $\mathbf{X}\in {\mathbb{C}^{t\times n}}$ and matrix $\mathbf{Y}\in {\mathbb{C}^{n\times s}}$,  we have $\mathrm{Rank}(\mathbf{X}\mathbf{Y})\geq\mathrm{Rank}(\mathbf{X})+\mathrm{Rank}(\mathbf{Y})-n$. Especially, if $\mathbf{X}\mathbf{Y}=\mathbf{0}$, then $\text{Rank}(\mathbf{X})+\text{Rank}(\mathbf{Y})\leq n$.

\emph{Lemma 2}: If $\mathbf{X}$ and $\mathbf{Y}$ are matrices with the same dimensions, it is always true that $\text{Rank}(\mathbf{X})+\text{Rank}(\mathbf{Y})\geq\text{Rank}(\mathbf{X}+\mathbf{Y})$.
\begin{IEEEproof}
\begin{eqnarray}
\text{Rank}(\mathbf{X})+\text{Rank}(\mathbf{Y})\ge \text{Rank}\left[ \begin{matrix}
   \mathbf{X}  \\
   \mathbf{Y}  \\
\end{matrix} \right]=\text{Rank}\left[ \begin{matrix}
   \mathbf{X}+\mathbf{Y}  \\
   \mathbf{Y}  \\
\end{matrix} \right]\\ \nonumber
\ge \text{Rank}\left[ \begin{matrix}
   \mathbf{X}+\mathbf{Y}  \\
   \mathbf{0}  \\
\end{matrix} \right]=\text{Rank}(\mathbf{X}+\mathbf{Y}).
\end{eqnarray}\label{prooflemma1}
\end{IEEEproof}
Now, we construct the lagrangian function of the problem (\ref{OP3.5}) with respect to $\mathbf{V}_{k,j}$, which is given by
\begin{eqnarray}
\mathcal{L}\left( {{\mathbf{V}}_{k,j}} \right)\!\!\!\!\!\!&=&\!\!\!\!\!\!\sum\limits_{k=1}^{K}{\sum\limits_{j=1}^{J}{\text{tr}\left( {{\mathbf{Z}}^{H}}{{\mathbf{H}}_{k}}{{\mathbf{V}}_{k,j}}\mathbf{H}_{k}^{H}\mathbf{Z} \right)}}\nonumber\\
&-&\sum\limits_{k=1}^{K}{\sum\limits_{j=1}^{J}{\frac{{{\lambda }_{k,j}}}{{{\gamma }_{k,j}}}\mathbf{u}_{k,j}^{H}{{\mathbf{H}}_{i}}{{\mathbf{V}}_{k,j}}\mathbf{H}_{k}^{H}{{\mathbf{u}}_{k,j}}}}\nonumber\\
&+&\sum\limits_{k=1}^{K}{\sum\limits_{j=1}^{J}{{{\lambda }_{k,j}} \sum\limits_{i=1,i\ne k}^{K}{\sum\limits_{m=1,m\ne j}^{J}{ \mathbf{u}_{k,j}^{H}{{\mathbf{H}}_{i}}{{\mathbf{V}}_{i,m}}\mathbf{H}_{i}^{H}{{\mathbf{u}}_{k,j}} }}}}\nonumber\\
&+&{{\lambda }_{k,j}} \left(\sum\limits_{i=1}^{K}{{{\left\| \mathbf{u}_{k,j}^{H}{{\mathbf{H}}_{i}}{{\mathbf{W}}_{i}} \right\|}^{2}}}+\sigma _{n}^{2}{{\left\| \mathbf{u}_{k,j}^{H} \right\|}^{2}} \right)\nonumber\\
&+&\sum\limits_{k=1}^{K}{{{\mu }_{k}}\left[ \sum\limits_{j=1}^{J}{\text{tr}\left( {{\mathbf{V}}_{k,j}} \right)+\left\| {{\mathbf{W}}_{k}} \right\|_{F}^{2}-{{P}_{\max ,k}}} \right]}\nonumber\\
&-&\sum\limits_{k=1}^{K}{\sum\limits_{j=1}^{J}{{\bm{\Psi }_{k,j}}{{\mathbf{V}}_{k,j}}}},
\end{eqnarray}
where $\lambda_{k,j}$, $\mu_k$ and ${{\bm{\Psi }_{k,j}}}, \forall k,j,$ are Lagrange multipliers of constraint (\ref{SINR1}), (\ref{OP3st1}) and (\ref{OP3st2}), respectively. Satisfied with the Slater's condition, we reveal the structure of the optimal $\mathbf{V}_{k,j}^{*}$ by exploiting the Karush-Kuhn-Tucher (KKT) conditions:
\begin{subequations}
\begin{equation}\label{p1}
\begin{split}
&\sum\limits_{i=1,i\ne k}^{K}{\sum\limits_{m=1,m\ne j}^{J}{\mathbf{u}_{k,j}^{H}{{\mathbf{H}}_{i}}\mathbf{V}_{i,m}^{*}\mathbf{H}_{i}^{H}{{\mathbf{u}}_{k,j}}}+\sum\limits_{i=1}^{K}{{{\left\| \mathbf{u}_{k,j}^{H}{{\mathbf{H}}_{i}}{{\mathbf{W}}_{i}} \right\|}^{2}}}}\\
&+\sigma _{n}^{2}{{\left\| \mathbf{u}_{k,j}^{H} \right\|}^{2}}\!-\!\frac{1}{{{\gamma }_{k,j}}}\left( \mathbf{u}_{k,j}^{H}{{\mathbf{H}}_{k}}\mathbf{V}_{k,j}^{*}\mathbf{H}_{k}^{H}{{\mathbf{u}}_{k,j}} \right)\!=\!0,
\end{split}
\end{equation}
\begin{equation}\label{p2}
  {{\nabla }_{\mathbf{V}_{k,j}^{*}}}\mathcal{L}=\mathbf{H}_{k}^{H}\mathbf{Z}{{\mathbf{Z}}^{H}}{{\mathbf{H}}_{k}}-\frac{\lambda _{k,j}^{*}}{{{\gamma }_{k,j}}}\mathbf{H}_{k}^{H}{{\mathbf{u}}_{k,j}}\mathbf{u}_{k,j}^{H}{{\mathbf{H}}_{k}}+\mu _{k}^{*}\mathbf{I}-{\bm{\Psi }_{k,j}^*}=0
\end{equation}
\begin{equation}\label{p3}
{\bm{\Psi }_{k,j}^{*}}{{\mathbf{V}}_{k,j}^{*}}=\mathbf{0},
\end{equation}
\begin{equation}\label{p4}
\lambda _{k,j}^{*}\ge 0,\mu _{k}^{*}\ge 0,\bm{\Psi} _{k,j}^{*}\succeq\mathbf{0}.
\end{equation}
\end{subequations}
From (\ref{p1}), it is known that ${{\mathbf{V}}_{k,j}^{*}}\neq\mathbf{0}$ due to $\sum\limits_{i=1}^{K}{{{\left\| \mathbf{u}_{k,j}^{H}{{\mathbf{H}}_{i}}{{\mathbf{W}}_{i}} \right\|}^{2}}+\sigma _{n}^{2}{{\left\| \mathbf{u}_{k,j}^{H} \right\|}^{2}}}>0$. In other words,
\begin{equation}\label{lp5.1}
  \text{Rank}(\mathbf{V}_{k,j}^{*})\ge 1.
\end{equation}
Then, applying Lemma 1 to (\ref{p3}), i.e., ${\bm{\Psi }_{k,j}^{*}}{{\mathbf{V}}_{k,j}^{*}}=\mathbf{0}$, we have
\begin{equation}\label{lp5.15}
 \text{Rank}(\bm{\Psi}_{k,j}^{*})+\text{Rank}(\mathbf{V}_{k,j}^{*})\le M.
\end{equation}
  Combined with (\ref{lp5.1}), we obtain
\begin{equation}\label{lp5.2}
  \text{Rank}(\bm{\Psi} _{k,j}^{*})\le M-1.
\end{equation}
Next, according to Lemma 2, we can deduce from (\ref{p2}) that
\begin{equation}\label{lp5.3}
  \text{Rank}(\bm{\Psi} _{k,j}^{*})+\text{Rank}(\bm{\Upsilon}_{k,j})\ge \text{Rank}(\mu _{k}^{*}\mathbf{I}),
\end{equation}
where $\bm{\Upsilon}_{k,j}=\mathbf{H}_{k}^{H}\left( \frac{\lambda _{k,j}^{*}}{{{\gamma }_{k,j}}}{{\mathbf{u}}_{k,j}}\mathbf{u}_{k,j}^{H}-\mathbf{Z}{{\mathbf{Z}}^{H}} \right){{\mathbf{H}}_{k}}$. Since $\bm{\Upsilon}_{k,j}\neq\mathbf{0}$, namely $\text{Rank}(\bm{\Upsilon}_{k,j})\geq1$, and $\text{Rank}(\mu _{k}^{*}\mathbf{I})=M$, we have
\begin{equation}\label{lp5.4}
    \text{Rank}(\bm{\Psi} _{k,j}^{*})\ge M-1.
\end{equation}
Combing (\ref{lp5.2}) and (\ref{lp5.4}), it is found that $ \text{Rank}(\bm{\Psi} _{k,j}^{*})= M-1$. Finally, substituting  $\text{Rank}(\bm{\Psi} _{k,j}^{*})= M-1$ into (\ref{lp5.15}), we obtain
\begin{equation}\label{lp5.5}
     \text{Rank}(\mathbf{V}_{k,j}^{*})\le 1.
\end{equation}
By (\ref{lp5.1}) and (\ref{lp5.5}), we can conclude that $\text{Rank}(\mathbf{V}_{k,j}^{*})=1$,  which means that the SDR processing for $\mathbf{V}_{k,j}=\mathbf{v}_{k,j}\mathbf{v}_{k,j}^{H}$ in the problem (\ref{OP3.5}) is tight. In other words, it makes up for the impact of dropping the rank-one constraint (\ref{OP3st3}) to the non-convex subproblem (\ref{OP3}).  The proof is completed.

\section{The Proof of Theorem 2}
For the received sensing signal $y_{k,j}^{'}$ at the BS, the MSE related to the $j$th sensing signal at the $k$th UE can be expressed as
\begin{eqnarray}\label{pmse1}
  \mathrm{MSE}_{k,j}&=&\mathbb{E}\left\{ \left( {{{y}}_{k,j}^{'}}-s_{k,j}^{'} \right){{\left( {{{y}}_{k,j}^{'}}-s_{k,j}^{'} \right)}^{H}} \right\}\nonumber\\
  &=&\sum\limits_{i=1}^{K}{\sum\limits_{m=1}^{J}{\mathbf{u}_{k,j}^{H}{{\mathbf{H}}_{i}}{{\mathbf{v}}_{i,m}}\mathbf{v}_{i,m}^{H}\mathbf{H}_{i}^{H}{{\mathbf{u}}_{k,j}}}}\nonumber\\
&+&\sum\limits_{i=1}^{K}{\mathbf{u}_{k,j}^{H}{{\mathbf{H}}_{i}}}{{\mathbf{W}}_{i}}\mathbf{W}_{i}^{H}\mathbf{H}_{i}^{H}{{\mathbf{u}}_{k,j}}+\sigma _{n}^{2}{{\left\| {{\mathbf{u}}_{k,j}} \right\|}^{2}}\nonumber\\
&-&\mathbf{u}_{k,j}^{H}{{\mathbf{H}}_{k}}{{\mathbf{v}}_{k,j}}-\mathbf{v}_{k,j}^{H}\mathbf{H}_{k}^{H}{{\mathbf{u}}_{k,j}}+1.
\end{eqnarray}
Based on the above equation, let us define ${\bm{\Omega }_{k,j}}=\sum\limits_{i=1}^{K}{\sum\limits_{m=1}^{J}{{{\mathbf{H}}_{i}}{{\mathbf{v}}_{i,m}}\mathbf{v}_{i,m}^{H}\mathbf{H}_{i}^{H}}}+\sum\limits_{i=1}^{K}{{{\mathbf{H}}_{i}}}{{\mathbf{W}}_{i}}\mathbf{W}_{i}^{H}\mathbf{H}_{i}^{H}+\sigma _{n}^{2}\mathbf{I}$, and then (\ref{pmse1}) can be rewritten as
\begin{eqnarray}
 \!\!\!\mathrm{MSE}_{k,j}\!\!\!\!&=&\!\!\!\!\mathbf{u}_{k,j}^{H}{\bm{\Omega}_{k,j}}{{\mathbf{u}}_{k,j}}-\mathbf{u}_{k,j}^{H}{{\mathbf{H}}_{k}}{{\mathbf{v}}_{k,j}}-\mathbf{v}_{k,j}^{H}\mathbf{H}_{k}^{H}{{\mathbf{u}}_{k,j}}+1\nonumber\\
  \!\!\!\!&=&\!\!\!\!\left( \mathbf{u}_{k,j}^{H}-\mathbf{v}_{k,j}^{H}\mathbf{H}_{k}^{H}\bm{\Omega} _{k,j}^{-1} \right){\bm{\Omega }_{k,j}}{{\left( \mathbf{u}_{k,j}^{H}-\mathbf{v}_{k,j}^{H}\mathbf{H}_{k}^{H}\bm{\Omega} _{k,j}^{-1} \right)}^{H}}\nonumber\\
  \!\!\!\!&+&\!\!\!\!1-\mathbf{v}_{k,j}^{H}\mathbf{H}_{k}^{H}\bm{\Omega} _{k,j}^{-H}{{\mathbf{H}}_{k}}{{\mathbf{v}}_{k,j}}.\label{pmse2}
\end{eqnarray}
It is found that the $\mathrm{MSE}_{k,j}$ is minimized only when ${{\mathbf{u}}_{k,j}}=\bm{\Omega} _{k,j}^{-H}{{\mathbf{H}}_{k}}{{\mathbf{v}}_{k,j}}$, namely employing the MMSE receiver. Since we adopt the MMSE receiver in the subproblem of optimizing receive beams, the MMSE associated with the sensing signal $s_{k,j}^{'}$ is given by
\begin{eqnarray}\label{pmse3}
{{e}_{k,j}}&=&1-\mathbf{v}_{k,j}^{H}\mathbf{H}_{k}^{H}\bm{\Omega} _{k,j}^{-H}{{\mathbf{H}}_{k}}{{\mathbf{v}}_{k,j}}\nonumber\\
&=&\frac{\bm{\Omega} _{k,j}^{H}-{{\mathbf{H}}_{k}}{{\mathbf{v}}_{k,j}}\mathbf{v}_{k,j}^{H}\mathbf{H}_{k}^{H}}{\bm{\Omega}_{k,j}^{H}}\nonumber\\
&=&\frac{\mathbf{u}_{k,j}^{H}\bm{\Omega} _{k,j}^{H}{{\mathbf{u}}_{k,j}}-\mathbf{u}_{k,j}^{H}{{\mathbf{H}}_{k}}{{\mathbf{v}}_{k,j}}\mathbf{v}_{k,j}^{H}\mathbf{H}_{k}^{H}{{\mathbf{u}}_{k,j}}}{\mathbf{u}_{k,j}^{H}\bm{\Omega} _{k,j}^{H}{{\mathbf{u}}_{k,j}}}\nonumber\\
&=&\frac{1}{1+{{\Gamma }_{k,j}}}.
\end{eqnarray}
The proof is completed.

\end{appendices}

\end{document}